\documentclass[conference]{IEEEtran}
\IEEEoverridecommandlockouts
\usepackage{cite}
\usepackage{amsmath,amssymb,amsfonts}
\usepackage{algorithmic}
\usepackage{graphicx}
\usepackage{multirow}
\usepackage{textcomp}
\usepackage{xcolor}
\def\BibTeX{{\rm B\kern-.05em{\sc i\kern-.025em b}\kern-.08em
    T\kern-.1667em\lower.7ex\hbox{E}\kern-.125emX}}
\begin{document}

\title{Real-Time Energy Measurement for Non-Intrusive Well-Being Monitoring of Elderly People -- a Case Study
\thanks{This work was supported by the National Centre for~Research and Development within the Smart Growth Operational Programme (agreement No.~POIR.01.02.00-00-0352/16-00).}
}

\author{\IEEEauthorblockN{~\\[-0.4ex]\large Mateusz Brzozowski \\[0.3ex]\normalsize}
\IEEEauthorblockA{OneMeter Sp. z o.o.\\ul. Dobrza\'{n}skiego 3\\
20-262 Lublin, Poland\\
Email: \tt Mateusz.Brzozowski@onemeter.com}
\and
\IEEEauthorblockN{~\\[-0.4ex]\large Artur Janicki\\[0.3ex]\normalsize}
\IEEEauthorblockA{Warsaw University of Technology\\
ul. Nowowiejska 15/19\\
00-665 Warsaw, Poland\\
Email: {\tt Artur.Janicki@pw.edu.pl}}
}

\maketitle

\begin{abstract}
This article presents a case study demonstrating a non-intrusive method for the well-being monitoring of elderly people. It is based on our real-time energy measurement system, which uses tiny beacons attached to electricity meters. Four participants aged 67--82 years took part in our study. We observed their electric power consumption for approx. a month, and then we analyzed them, taking into account the participants' notes on their activities. We created typical daily usage profiles for each participant and used anomaly detection to find unusual energy consumption.  We found out that real-time energy measurement can give significant insight into someone's daily activities and, consequently, bring invaluable information to caregivers about the well-being of an elderly person, while being discreet and entirely non-intrusive.
\end{abstract}

\begin{IEEEkeywords}
energy measurement, smart metering, assistive technology, assisted ambient living, anomaly detection, non-intrusive monitoring
\end{IEEEkeywords}

\section{Introduction}
Our societies are getting older. The growing number of older people requires, among other things, effective yet discreet methods of monitoring the well-being of the elderly by their caregivers, such as children or other relatives. In this article, we present a method for non-invasive remote monitoring of a person's well-being at their residence based on real-time electricity measurement. Our method is based on the electric energy measurement system presented in~\cite{Brzozowski2019Enhancing}. It was originally proposed to speed up the development of smart metering. However, here, we will present its application to assistive technologies.

The proposed monitoring method does not require any interference with installations in the user's residence, nor does it involve using any wearable devices, such as smartwatches or wristbands. It does not require the installation of a camera or microphone at the user's residence, which could be considered invasive to privacy. The proposed method is suitable for use with different types of electricity meters, which may or may not be smart. It is sufficient that they are equipped with an optical port, a flashing diode, or another port for communication (e.g., galvanic). 

In this paper, in Section~\ref{sec:SOTA}, we will briefly present existing solutions for monitoring elderly people in their homes. Next, in Section~\ref{sec:Method}, we will describe the proposed method for non-invasive monitoring of elderly people using energy measurement based on our approach. In Section~\ref{sec:Experiments}), we will present the setup of our study, followed by initial results, presented in~Section~\ref{sec:Results}. Finally, we will conclude in Section~\ref{sec:Conclusions} with a plan for the future of our work.

\begin{figure}[!t]
\centering
\includegraphics[width=0.7\linewidth]{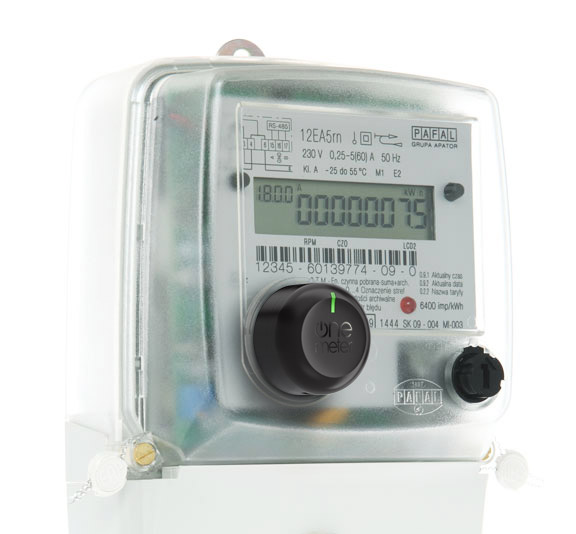}
\caption{OneMeter beacon with optical sensor attached to IEC 62056-21 port of electricity meter.}
\label{fig:beacon}
\end{figure}


\section{Monitoring Elderly People}
\label{sec:SOTA}

Various methods of remote monitoring of human well-being are known. For example, video and voice monitoring systems are used, which involve installing devices such as motion sensors, cameras, and microphones at the residence of the person being monitored, forming the so-called assisted living system (ALS)~\cite{Daher2018Multisensory, Gingras2020Toward}. These devices are designed to check how the person moves around the apartment, such as whether the person gets out of bed, opens the medicine cabinet, walks from the room to the kitchen, etc. The microphone or microphone arrays, on the other hand, are designed to record disturbing noises that, for example, may indicate a fall~\cite{Li2012Microphone}. Another way is to equip the monitored person with various types of sensors, such as a wristband, smartwatch, or chest-mounted sensors~\cite{Gupta2007Wireless, Ianculescu2021Fostering}. These monitor the monitored person's vitals, such as pulse or oxygen saturation. 

Some systems analyze electricity consumption by inserting an appropriate probe into the user's electrical system, which measures energy independently of the measurement that the electricity supplier makes. This is called non-intrusive load monitoring (NILM) and has been described, e.g., in~\cite{Hernandez2022Evaluating, Hernandez2023Estimating} for helping patients with Alzheimer's disease. 



However, the methods of monitoring human well-being discussed above have several disadvantages. For example, they require the installation of devices that can invade the privacy of the monitored person, such as sensors, cameras, and microphones, making them often unacceptable to the monitored person. Other methods require the monitored person to wear devices such as wristbands or other wearables, which is also inconvenient and, therefore, often rejected. In addition, the monitored person may remove their wearables (e.g., before bathing) and forget to put them back on. These devices also require regular charging. Once they are discharged, they stop working, which can cause interruptions in monitoring and generating false alarms. Finally, other existing methods require interference with the electrical system at the residence of the monitored person, which is costly and requires employing qualified staff, and, as a consequence, it may also not be accepted by the monitored person. To sum up, contrary to our solution, none of the above-described approaches is truly non-invasive.

\begin{table*}[htbp]
\caption{Study participants and electrical equipment they used}
\begin{center}
\begin{tabular}{|c|c|c|c|c|c|c|c|c|c|c|c|c|c|}
\hline
\multirow{3}{*}{\textbf{User}} &\multirow{3}{*}{\textbf{Age}} &\multirow{3}{*}{\textbf{Sex}} &\multicolumn{2}{|c|}{\textbf{Lighting}} &\multicolumn{9}{|c|}{\textbf{Electrical devices used}} \\
\cline{4-14} 
 & & & \multirow{2}{*}{\textbf{\textit{Regular}}} &\multirow{2}{*}{\textbf{\textit{LED}}} & \multirow{2}{*}{\textbf{\textit{Fridge}}} & \multirow{2}{*}{\textbf{\textit{Kettle}}} & \multirow{2}{*}{\textbf{\textit{Oven}}}& \textbf{\textit{Dish-}}&  \textbf{\textit{Hair-}} &  \textbf{\textit{Washing}}& \multirow{2}{*}{\textbf{\textit{Iron}}} & \multirow{2}{*}{\textbf{\textit{TV}}} & \multirow{2}{*}{\textbf{\textit{Other}}}\\
 & & &  & &  &  & & \textbf{\textit{washer}}&  \textbf{\textit{dryer}} &  \textbf{\textit{machine}} & &  & \\
\hline
S1 & 82 & F & - & x & x & x & x & x & x & x & x & x & -  \\
S2 & 76 & M & x & x & x & x & x & x & x & x & x & x & Alarm  \\
S3 & 77 & F & - & x & x & - & - & - & - & x & - & x & - \\
S4 & 67 & F & - & x & x & x & x & - & x & x & x & x & A/C \\
\hline
\end{tabular}
\label{tab1}
\end{center}
\end{table*}

\begin{figure*}[!ht]
\includegraphics[width=\linewidth]{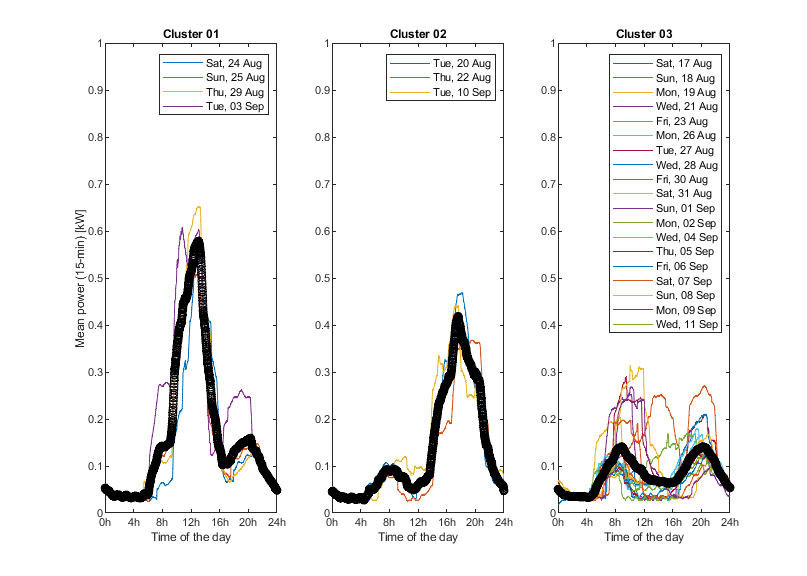}
\caption{15-min mean power profiles for clustered into three clusters for monitored user S4. The thick black line shows the mean profiles for each cluster.}
\label{fig:Clusters}
\end{figure*}

\section{Proposed method}
\label{sec:Method}

\begin{figure*}[!ht]
\includegraphics{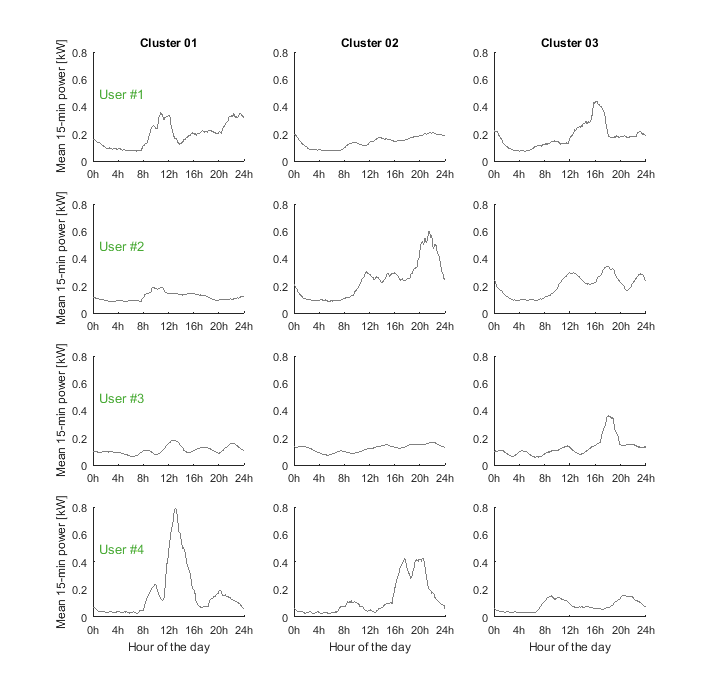}
\caption{Mean 15-min mean power profiles for three clusters identified per each monitored user.}
\label{fig:Clusters4users}
\end{figure*}

The idea of our solution is based on using a small, energy-efficient wireless beacon with optical sensors attached to an electricity meter (see Figure~\ref{fig:beacon}). A small bottle cap-shaped beacon of 32~mm diameter (compatible with the IEC~62056-21 standard) was designed, equipped with an optical sensor, LED diode, Nordic Semiconductor's processor nRF51, flash memory, Bluetooth Low Energy (BLE) radio components, and a 3.0V battery (CR2032). The beacon is attached magnetically to an electronic meter equipped with an optical port. The optical sensor is designed with a miniature silicon photodiode of high radiant sensitivity and a low-power comparator. The optical sensor, together with the IR LED diode, are able to set up communication with a meter using the IEC~62056-21 (old: IEC~1107) protocol. The amount of measurement data acquired from the meter depends on the meter's model -- some of the meters present only the absolute active energy, while the others allow the readout of more detailed information, such as positive and negative active energy, or reactive energy. 

The processor is programmed so that the beacon performs a readout of the meter every 15~min and stores the metering data in the flash memory. The BLE component allows other BLE devices to connect to the beacon to download metering data or to transmit the readout in real-time. In the original version~\cite{Brzozowski2019Enhancing}, we proposed employing either smartphones or dedicated gateways to transfer measurement data to the cloud. For real-time transmission, the preferable option would be using gateways, e.g., based on the LoRaWAN protocol, as described in our another study~\cite{Marcul2024System}.

Energy measurement data are then securely transmitted to the data platform. It provides gathering, analysis, and visualizations of the collected metering data. 
It can also allow performing analyses that can be used for non-intrusive monitoring of people's well-being. In this case study, we will show that the measurement data can be used to verify if a user follows a regular daily routine or, using anomaly detection, if the energy consumption is somehow anomalous, thus implying potential problems with the user's well-being.

\section{Experiments}
\label{sec:Experiments}

\begin{figure*}[!ht]
\includegraphics[width=\linewidth]{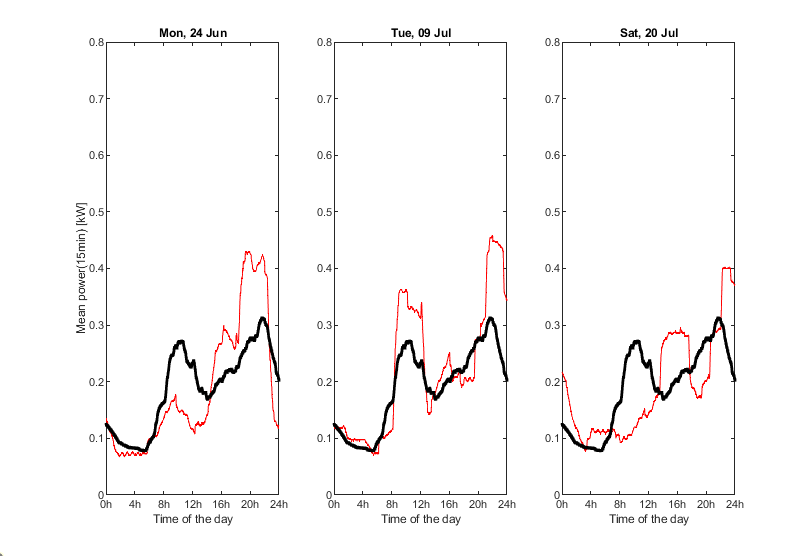}
\caption{Three anomalous days detected for user S1. Black lines denote the closest mean usage profile; red lines denote the usage from the anomalous days.}
\label{fig:Anomalies}
\end{figure*}

\subsection{Participants}
Four participants (named S1--S4), aged 67--82 yrs, took part in our study. They had the energy monitoring system installed for approx. one month, between June and September, in Warsaw, Poland. Their households were screened for electrical equipment being used. The participants were also asked to make notes on their home activities and their own well-being. Each of the participants signed a consent to take part in this study.

Each of the participants lived alone in their homes. Each household was lit with LED lighting; in addition, participant S2 used traditional electric bulbs. All households were equipped with a fridge, a TV set, and a washing machine. Half of the participants used a dishwasher. Even though all participants had hair dryers, S3 never used them, as she let her hair dry naturally. Similarly, she never ironed her clothes, but she let them dry naturally. She also used gas to cook and make tea or coffee. Participant S3 had an alarm system at home. Participant S4 had a portable air-conditioning system, which she used on hot days. Noteworthy, in none of the households in this study, the electricity was used for house heating or domestic water heating.

The participants also differed in terms of lifestyles. Participants S1 and S2 occasionally received guests, and the others did not. Participant S2 traveled and was absent for a couple of days. Participant S2 thoroughly cleaned the house once a week, while the others did it irregularly. Participant S4 often used her washing machine. All the above-presented information will be needed when analyzing the mean power profiles.

\subsection{Data analysis}
We analyzed the collected energy measurement data by calculating the 15-min mean power value throughout the study's timespan. For each participant, we analyzed daily profiles of these values. Next, using a clustering algorithm (in our case, the $k$-Means algorithm, originating from the signal processing domain~\cite{Lloyd1982Least}), we grouped the daily profiles into $k$ clusters, changing $k$ in the range from $1$ to $6$.

Next, we employed an anomaly detection algorithm by identifying time series (here: daily profiles), which were the furthest from the cluster's centroids. We compared this information with the participants' notes. The results are presented and described in detail in the next section.

\section{Results}
\label{sec:Results}

15-minute mean power profiles for each user have been collected and analyzed. Fig.~\ref{fig:Clusters} shows the result of grouping them into three clusters for one of the users, S1. The figure displays daily profiles and the average profiles in each of the clusters. As one can see, the k-Means algorithm was able to identify three typical daily routines of the S4 user:
\begin{itemize}
    \item Cluster 1: with increased power consumption around noon. The S4 user reported the following activities for these hours on the days attributed to these clusters: cooking lunch on an electric oven, cleaning the house with a vacuum cleaner, and doing laundry.
    \item Cluster 2: with increased power consumption in the late afternoon. Activities recorded: baking a cake for a name day party, running an air-conditioning;
    \item Cluster 3: the most typical routine: slightly increased power consumption in the morning, around 8 a.m. (using a hair-dryer, boiling water for tea/coffee, etc.), and similarly slightly increased power consumption in the evening, around 8--9 p.m. (watching TV, using lighting).
\end{itemize}
\noindent At night, the power consumption was minimal (probably only the fridge consumed power, plus devices on stand-by, like a TV or phone charger).

As one can see in~Fig.~\ref{fig:Clusters4users}, similarly, such clusters were created for all four participants of this study. Usually, one of these clusters was the most populated, as it contained ``the most typical'' daily routine (as Cluster 3 in the above example for user S4), with preparing meals, watching TV, running a dishwasher, etc. The other clusters, according to the participants' reports, could be attributed to the days when the participants received guests, did more cooking (e.g., baking), did laundry, or cleaned the house with vacuum cleaners. 

We researched various numbers of clusters (from 1 to 6), but it turned out that three clusters were sufficient to cover the variability of daily routines for the study participants and the number of days being analyzed.
One exception here was the S3 user, for whom the mean power profiles in each of the three clusters were very similar, indicating that a lower number of clusters (two, or even one) might be sufficient. In this case, the most populated was Cluster~1, showing very low variability within a day. It turned out that this user rarely used any electric devices, apart from LED lighting and energy-saving TV. The S3 user used gas for cooking, rarely ironed clothes, and always let her hair dry naturally. Therefore, the daily fluctuations in power consumption were minimal, and the possibility of monitoring the user's well-being in such a case would be very limited.

Having grouped the users' mean power profiles into clusters, we tried to use them to identify anomalous days to see if changes in user behavior could be captured by monitoring electric energy consumption. To do so, we found the days with the mean power profiles most distant (in the Euclidean sense) from the closest average profile. Fig.~\ref{fig:Anomalies} shows such profiles for user S1. According to the S1 participant, the identified days corresponded to the days with extensive evening baking accompanied with washing dishes in a dishwasher (24 Jun), baking a cake in the morning (9 Jul), or being absent in the morning (20 Jul) and shifting morning activities (preparing a meal, washing hair) a couple of hours later. As one can see, observing energy consumption can give significant insight into someone's daily activities and, as a consequence, bring invaluable information to caregivers about the well-being of an elderly person, or another person who lives on their own but needs discreet, non-intrusive monitoring.

\section{Conclusions and Future Work}
\label{sec:Conclusions}

The proposed method offers an entirely non-intrusive method of monitoring elderly people or any other people who need to be discreetly supervised, such as by their relatives or other caregivers.
The method works best if the supervised person uses more electrical devices, such as an electric kettle or an oven. Low-power devices, such as radio receivers, may not be readily visible in our measurements. If a person uses fewer electrical devices, information about, e.g., someone's reduced activity may be observed with some delay. This delay can depend on the time of year, which affects the usage of electric lighting at home. 

It should be noted that information about someone's energy consumption can pose a threat to their privacy. This means that the energy measurement data should be securely transmitted, stored, and analyzed. It is important to ensure that a person who is monitored consents to the monitoring. 

In the future, observing a larger group of participants over a much longer time span would be very advantageous. It would allow for the taking into account of changes in daylight cycles, which influence energy consumption. It would also allow us to observe long-term changes in the users' behavior.

\bibliographystyle{IEEEtran}
\bibliography{seniorEnergy}

\end{document}